# Dipolar order mapping based on spin-lock magnetic resonance imaging

Zijian Gao[1], Qianxue Shan[1], Ziqin Zhou[1,2], Ziqiang Yu[1], Weitian Chen[1,*]

1.Department of Imaging and Interventional Radiology, Faculty of Medicine, The Chinese University of Hong Kong, Hong Kong.

2.MR Research Collaboration, Siemens Heathineers Ltd., Hong Kong.

*Corresponding author:

Weitian Chen

Department of Imaging & Interventional Radiology

The Chinese University of Hong Kong

Shatin, Hong Kong, SAR

(852)-3505-1036

Email: wtchen@cuhk.edu.hk

# Abstract

## Purpose

Inhomogeneous magnetization transfer (ihMT) effect reflects dipolar order with a dipolar relaxation time ($T_{1D}$), specific to motion-restricted macromolecules. We aim to quantify $T_{1D}$ using a spin-lock MRI technique.

## Methods

In the proposed method, we introduce a $T_{1D}$-specific ratio, denoted as $RATIO_{dosl}$. This ratio is derived from the distinct relaxation rate $R_{dosl}$, calculated as the difference between dual-frequency $R_{1\rho}^{dual}$ relaxation and single-frequency $R_{1\rho}^{single}$ relaxation measurements. A novel rotary-echo spin-lock sequence was developed to enable dual-frequency spin-lock acquisition. We established a framework to estimate $T_{1D}$, as well as the macromolecular pool fraction (MPF) map. The proposed approach was validated via numerical simulations, phantom studies, and demonstrated *in vivo* in human white matter.

## Results

Simulations revealed the high sensitivity of $RATIO_{dosl}$ to $T_{1D}$, and substantiated the accuracy and robustness of the proposed methods. Phantom experiments demonstrated robust ihMT contrast and confirmed the capability of $T_{1D}$ quantification via $RATIO_{dosl}$. *In vivo* studies supported the clinical viability of this approach, achieving simultaneous $T_{1D}$ and MPF mapping using only three spin-lock prepared images. Across ten healthy volunteers, the mean white matter $T_{1D}$ ranged from approximately 3.70 to 4.80 ms.

## Conclusion

We propose a novel method for $T_{1D}$ quantification based on spin-lock MRI. By requiring only three contrast-prepared images, this technique provides a promising pathway for robust, rapid, and simultaneous $T_{1D}$ and MPF quantification with fewer confounds.

# 1.Introduction

In ordered tissues containing motion-restricted macromolecules, such as myelin, dipolar order exists alongside Zeeman order. This phenomenon manifests in magnetization transfer (MT) experiments as an asymmetric spectrum following single-frequency saturation, in contrast to the symmetric spectrum observed under dual-frequency saturation. Known as inhomogeneous magnetization transfer (ihMT), this effect can be quantified by the dipolar relaxation time, $T_{1D}$, and is highly sensitive to tissue microstructure[1–5]. Consequently, ihMT offers a promising approach for assessing microstructural integrity, particularly brain myelination[6,7]. It holds significant potential for elucidating disease pathophysiology (e.g., in multiple sclerosis) and providing valuable outcome measures to monitor neuroprotection and repair in clinical trials[8–10].

The two-pool model is commonly used to analyze the MT effect, while Provotorov theory, as formulated by Goldman for the dipolar order effect[1], indicates that the MT pool can be subdivided into a Zeeman reservoir and a dipolar reservoir. The parameter $\beta$, proportional to the inverse spin temperature of the dipolar reservoir, is incorporated into the two-pool model[2,3,11]. To measure dipolar order under *in vivo* conditions, Varma et. al introduced ihMT ratio (ihMTR) for *in vivo* experiments to indicate the ihMT effect via a subtraction experiment between images acquired with single frequency saturation and dual frequency saturation[4]. They further proposed $T_{1D}$ quantification using multiple ihMTR images with varied switch times during dual-frequency saturation[6]. However, this approach requires long scan time to acquire sufficient ihMTR images (e.g. eight ihMTR images with different switch times[6]). Prevost et al. proposed an optimized ihMT acquisition strategy based on $T_{1D}$ filtering effect[12], which enables the isolation of short- and long- $T_{1D}$ components by manipulating switch times[13,14]. A pseudo-quantitative ihMT (qihMT) method has also been developed based on inverse subtraction metrics[9,15]. This approach enables fast acquisition and high reproducibility with fewer confounding factors, and it has been frequently used in human studies. [16–18]. Malik et al. proposed a rapid steady-state pulse sequence for ihMT contrast using multiband radiofrequency pulses, which is used to simultaneously perform off-resonance saturation and on-resonance excitation[19]. Nevertheless, these techniques remain semi-quantitative without directly quantifying dipolar order. Recently, West et al proposed a MR fingerprinting method to quantify $T_{1D}$, macromolecular proton fraction (MPF) and $T_1$ of free water pool[20]. Despite these challenges, pursuing $T_{1D}$ quantification to assess the microstructural integrity of myelin with minimized confounds and clinically feasible scan time remains highly valuable[8,21,22].

Recently, an off-resonance spin-lock quantitative MT approach (MPF-SL) was proposed to mitigate free water pool contributions[23–25], suppress residual dipolar coupling (RDC) effects from motion-restricted water molecules[26,27], and enable rapid measurement of the macromolecular proton fraction (MPF)[28], offering strong potential for clinical application. In MPF-SL, appropriately designed off-resonance spin-lock acquisitions are used to remove the relaxation contribution associated with the free water pool (the free water $T_1$ and $T_2$) from the rotating frame relaxation rate ($R_{1\rho}$). This yields an MT-specific relaxation rate, $R_{mpfsl}$, from which MPF is derived. By isolating the macromolecular contribution, MPF-SL eliminates the need for prior knowledge or estimation of the free water $T_1$ and $T_2$, thereby significantly reducing the complexity of MPF quantification.

In this study, we extend this approach to rapidly quantify the dipolar relaxation time, $T_{1D}$. With our proposed method, both the MPF and $T_{1D}$ can be acquired in a single fast scan, providing complementary information for tissue assessment. To probe the ihMT effect, we utilize a chain of rotary echo spin-lock radiofrequency (RF) pulses with variable switch times. These pulses alternate between positive and negative resonance frequency offsets, synchronized with RF phase alterations, to maintain magnetization spin-locking throughout the pulse duration. These spin-lock acquisitions eliminate free water contributions from the signal model, thereby substantially simplifying the calculation of $T_{1D}$. Furthermore, MPF is quantified from the same dataset without requiring additional scans. We demonstrate the efficacy of this method using numerical simulations, phantom experiments, and in vivo measurements.

# 2.Theory

In two-pool model for MT, tissue magnetization is commonly divided into the water pool ($M_x^a, M_y^a, M_z^a$) and the MT pool ($M_z^b$)[29]. Based on Provotorov theory as formulated by Goldman[1], the MT pool is modified to include a Zeeman reservoir and a dipolar reservoir. The parameter $\beta$, which is proportional to the inverse spin temperature of the dipolar reservoir, is incorporated into the magnetization vector[2,3,11]

$$\vec{M} = (M_{ax}, M_{ay}, M_{az}, M_{bz}, \beta)^T \quad , \tag{1}$$

which follows:

$$\frac{d}{dt}\vec{M} = A \cdot \vec{M} + \vec{C} \quad . \tag{2}$$

Following the notation by Zaiss et.al[30] and incorporating dipolar order[11], we present A as a 5 x 5 system matrix:

$$A = \begin{pmatrix} -R_{2a} & -\Delta\omega & 0 & 0 & 0 \\ +\Delta\omega & -R_{2a} & +\omega_1 & 0 & 0 \\ 0 & -\omega_1 & -R_{1a} - k_{ab} & k_{ba} & 0 \\ 0 & 0 & k_{ab} & -R_{1b} - R_{rfb} - k_{ba} & R_{rfb}\Delta\omega \\ 0 & 0 & 0 & R_{rfb}\frac{\Delta\omega}{D^2} & -(\frac{1}{T_{1D}} + R_{rfb}(\frac{\Delta\omega}{D})^2) \end{pmatrix} \tag{3}$$

and $\vec{C}$ is a constant vector:

$$\vec{C} = (0,0,R_{1\mathrm{a}}M_{0\mathrm{a}},R_{1\mathrm{b}}M_{0\mathrm{b}},0)^{\mathrm{T}} \tag{4}$$

where $T_{1D}$ is dipolar relaxation time. $R_{2a}$ and $R_{1a}$ are the transverse and longitudinal relaxation rates for the water pool, respectively. $R_{1b}$ is the longitudinal relaxation rate for the MT pool. $R_{rfb} = \omega_1^2 \pi g(T_{2b},\Delta\omega)$, where $g(T_{2b},\Delta\omega)$ denotes the absorption lineshape of the MT pool, and we use a super-Lorentzian lineshape model in this study[3]. $T_{2b}$ is a parameter of the MT pool that characterizes its absorption lineshape. $M_{0\mathrm{a}}$ and $M_{0\mathrm{b}}$ denote the equilibrium magnetizations of the water and MT pools, respectively. $\Delta\omega$ is the resonance frequency offset (FO) and $\omega_1$ is the frequency of spin-lock (FSL). $k_{ab}$ and $k_{ba}$ are the exchange rates between the water pool and the MT pool. D is associated with the local dipolar field (in angular frequency units)[3], which approximately equals to $\frac{1}{T_{2b}\sqrt{15}}$. In addition, we assume the fraction of dipolar order $f_D$=1[3,11].

When dual-frequency RF is applied with simultaneous irradiation at positive and negative frequency, the term proportional to $\Delta\omega$ cancels in Eq.3. It indicates the contribution from dipolar reservoir can be removed using dual-frequency RF irradiation. Consequently, $A_{dual}$ simplifies to:

$$A_{dual} = \begin{pmatrix} -R_{2a} & -\Delta\omega & 0 & 0 \\ +\Delta\omega & -R_{2a} & +\omega_1 & 0 \\ 0 & -\omega_1 & -R_{1a} - k_{ab} & +k_{ba} \\ 0 & 0 & +k_{ab} & -R_{1b} - R_{rfb} - k_{ba} \end{pmatrix} \tag{5}$$

In the rotating frame, the relaxation rates $R_{1\rho}$ measured under single-frequency spin-lock irradiation ($R_{1\rho}^{\mathrm{single}}$, with $\Delta\omega^s$ and $\omega_1^s$) and dual-frequency spin-lock irradiation ($R_{1\rho}^{\mathrm{dual}}$, with $\Delta\omega^d$ and $\omega_1^d$) are primarily governed by the least negative eigenvalue of $A$ and $A_{dual}$ [30,31], respectively:

$$R_{1\rho}^{single} = R_W(\Delta\omega^s, {\omega_1}^s) + R_{MT}^s(\Delta\omega^s, {\omega_1}^s) \tag{6}$$

and

$$R_{1\rho}^{dual} = R_W\left(\Delta\omega^d, {\omega_1}^d\right) + R_{MT}^d\left(\Delta\omega^d, {\omega_1}^d\right) \tag{7}$$

Here, $R_W$ represents the effective relaxation rate of the water pool, while $R_{MT}^s$ and $R_{MT}^d$ denote the relaxation rates associated with the MT pool under single-frequency and dual-frequency spin-lock irradiation, respectively. Notably, $R_{1\rho}^{single}$ includes a contribution from dipolar order, whereas this contribution is suppressed in $R_{1\rho}^{\mathrm{dual}}$. Under the conditions $\Delta\omega^d \gg {\omega_1}^d$ and $\Delta\omega^s \gg {\omega_1}^s$, the influence of chemical exchange is negligible. When single-frequency and dual-frequency spin-lock are applied with the same direction of spin-lock field (i.e., $\frac{\Delta\omega^d}{\omega_1^d} = \frac{\Delta\omega^s}{\omega_1^s}$) , the water pool contribution $R_W$ can be removed by calculating the difference between $R_{1\rho}^{single}$ and $R_{1\rho}^{dual}$. Consequently, we defined a distinct relaxation rate $R_{dosl}$ associated with the ihMT effect (see Supporting Information 1 for details):

$$
\begin{aligned}
R_{dosl} &= R_{1\rho}^{dual} - R_{1\rho}^{single} = R_{MT}^d - R_{MT}^s \\
&= f_b k_{ba}\left[\frac{R_{rfb}^d}{k_{ba}(f_b+1)+R_{rfb}^d} - \frac{R_{rfb}^s}{k_{ba}(f_b+1)(R_{rfb}^s\left(\frac{\Delta\omega^s}{D}\right)^2 T_{1D}+1)+R_{rfb}^s}\right]
\end{aligned}
\qquad (8)
$$

Here, $R_{rfb}^s$ and $R_{rfb}^d$ denote the lineshape under single-frequency and dual-frequency spin-lock conditions, respectively. By definition, $k_{ba} = R(1-f_b)$, where $R$ is the exchange rate. Since $f_b$ typically lies in the range 0.08-0.16, implying $f_b^2 \ll 1$, we have $k_{ba}(f_b+1) = R(1-f_b^2) \approx R$. Thus Eq. (8) simplifies to:

$$
\begin{aligned}
R_{dosl} &= R_{1\rho}^{dual} - R_{1\rho}^{single} = R_{MT}^d - R_{MT}^s \\
&= f_b R(1-f_b)\left[\frac{R_{rfb}^d}{R+R_{rfb}^d} - \frac{R_{rfb}^s}{R(R_{rfb}^s\left(\frac{\Delta\omega^s}{D}\right)^2 T_{1D}+1)+R_{rfb}^s}\right]
\end{aligned}
\qquad (9)
$$

To further improve the robustness of the calculation of $T_{1D}$, we introduce a variable, $RATIO_{dosl}$, obtained by measuring the ratio of $R_{dosl}$ at two different spin-lock conditions (denoted as $R_{dosl,1}$ and $R_{dosl,2}$) with distinct combinations of $\Delta\omega$ and $\omega_1$. We impose the constraints: $\frac{\Delta\omega^{d(1)}}{\omega_1^{d(1)}} = \frac{\Delta\omega^{d(2)}}{\omega_1^{d(2)}} = \frac{\Delta\omega^{s(1)}}{\omega_1^{s(1)}}$, $\Delta\omega^{d(1)} = \Delta\omega^{s(1)}$, and $\frac{\Delta\omega^{d(1)}}{\Delta\omega^{d(2)}} = N\ (N > 1)$. Consequently, $RATIO_{dosl}$ is calculated as follows:

$$RATIO_{dosl} = \frac{R_{dosl,1}}{R_{dosl,2}} = \frac{R_{1\rho}^{dual\,(1)} - R_{1\rho}^{single\,(1)}}{R_{1\rho}^{dual\,(2)} - R_{1\rho}^{single\,(1)}}$$

$$= \left[ \frac{R + R_{rfb}^{d(2)}}{R + R_{rfb}^{d(1)}} * \frac{R_{rfb}^{d(1)} R_{rfb}^{s(1)} \left(\frac{\Delta\omega^{s(1)}}{D}\right)^2 T_{1D}}{R_{rfb}^{d(2)} \left( R_{rfb}^{s(1)} \left(\frac{\Delta\omega^{s(1)}}{D}\right)^2 T_{1D} + 1 \right) - R_{rfb}^{s(1)}} \right]$$

(10)

In quantitative MT, the parameter $R$ and the lineshape parameter $T_{2b}$ are often treated as constants in human studies[32]. Under this assumption, $RATIO_{dosl}$ is primarily associated with the dipolar relaxation time $T_{1D}$, allowing $T_{1D}$ to be determined conveniently from $RATIO_{dosl}$. Notably, MPF can be calculated from $R_{1\rho}^{\mathrm{dual}(1)}$and $R_{1\rho}^{\mathrm{dual}(2)}$ following the method of Hou et.al[23]. This approach eliminates the confounding influence of the dipolar order term on the MPF calculation without requiring additional data acquisition.

# 3.Method

## 3.1 Acquisition scheme

In existing ihMT acquisition schemes, dual-frequency saturation can be applied using rapid alternation at positive and negative frequency on a minimal timescale[6]. Similarly, we propose using a chain of rotary echo spin-lock RF pulses to probe the ihMT effect. In this implementation, the spin-lock RF pulses alternate between positive and negative resonance frequency offsets with a switching time, $T_s$. To maintain magnetization spin-locking throughout the pulse duration, the RF phase is alternated in synchronization with the frequency offsets. By the $T_{1D}$ filtering effect[13,14], dual-frequency spin-lock is achieved when $T_s$ is shorter than the tissue $T_{1D}$ (Figure 1 (a)). In contrast, single-frequency spin-lock is implemented with a long $T_s$ that considerably exceeds the tissue $T_{1D}$ (Figure 1 (b)). This modified rotary-echo spin-lock RF pulse sequence provides a practical method to acquire $R_{1\rho}^{single}$ and $R_{1\rho}^{dual}$.

For *in vivo* experiment, directly measuring $R_{1\rho}^{single}$ and $R_{1\rho}^{dual}$ can be challenging. It requires multiple spin-lock prepared images with sufficiently long time of spin-lock (TSL) for robust quantification which is constrained by specific absorption rate (SAR) and RF hardware limitations. Following the approach reported by Hou et. al [23,28], we can collect data and calculate the difference of $R_{1\rho}^{single}$ and $R_{1\rho}^{dual}$ directly, rather than measuring them individually. Specifically, $R_{dosl} = R_{1\rho}^{dual} - R_{1\rho}^{single}$ can be obtained from two spin-lock prepared images using the fast acquisition strategy[28]. Consequently, $RATIO_{dosl}$ can be calculated using a total of only three spin-lock prepared images:

$$RATIO_{dosl} = \frac{R_{dosl,1}}{R_{dosl,2}} = \frac{R_{1\rho}^{dual\,(1)} - R_{1\rho}^{single\,(1)}}{R_{1\rho}^{dual\,(2)} - R_{1\rho}^{single\,(1)}}$$

$$= \frac{-\log\left(\frac{M_d^{(1)} - M_{d,Tog}^{(1)}}{M_s^{(1)} - M_{s,Tog}^{(1)}}\right)/TSL}{-\log\left(\frac{M_d^{(2)} - M_{d,Tog}^{(2)}}{M_s^{(1)} - M_{s,Tog}^{(1)}}\right)/TSL}$$

$$\approx \frac{-\log\left(\frac{M_d^{(1)}}{M_s^{(1)}}\right)/TSL}{-\log\left(\frac{M_d^{(2)}}{M_s^{(1)}}\right)/TSL} \tag{11}$$

Here, $M_d^{(1)}$ and $M_d^{(2)}$ represent the dual-frequency spin-lock prepared images corresponding to relaxation rates $R_{1\rho}^{dual\,(1)}$ and $R_{1\rho}^{dual\,(2)}$, respectively. $M_s^{(1)}$ is the single-frequency spin-lock prepared image associated with $R_{1\rho}^{single\,(1)}$. The subscript 'Tog' indicates the spin-lock prepared images acquired with manipulation of the initial magnetization. This images can be used to remove the influence from steady-state magnetization[23]. In the brain applications, the acquisition of this images can be omitted according to the theory described in the fast acquisition strategy by Hou et al[28].

## 3.2 Calculation of $T_{1D}$

The workflow for calculating $T_{1D}$ consists of two steps, as illustrated in Figure 2. In the first step, $RATIO_{dosl}$ is computed from $R_{dosl,1}$ and $R_{dosl,2}$ using three off-resonance spin-lock prepared images according to Eq. 11. In the second step, the $T_{1D}$ map is estimated from $RATIO_{dosl}$. While multiple numerical approaches exist for this estimation, this study investigates both an analytical estimation and a dictionary-matching approach. The analytical estimation determines $T_{1D}$ directly by solving Eq. 10 using experimentally observed $RATIO_{dosl}^{acq}$. In contrast, the dictionary-matching approach generates a $RATIO_{dosl}$ dictionary over a predefined range of $T_{1D}$ values; $T_{1D}$ is then determined by matching the measured $RATIO_{dosl}$ to the corresponding entry in the dictionary. Under fixed spin-lock pulse settings and tissue parameters of the two-pool model, $RATIO_{dosl}$ is a one-to-one function of $T_{1D}$. Accordingly, for each candidate $T_{1D}$, three magnetizations ($M_d^{(1)}$, $M_d^{(2)}$ and $M_s^{(1)}$) are simulated using Bloch-McConnell-Provotorov equation (Eq.1-5) following the modified rotary-echo spin-lock RF pulses (Figure 1), and the corresponding $RATIO_{dosl}$ is then derived via Eq.11. The spin-lock parameters are fixed to match the practical MRI acquisition setup, and tissue parameters of the water pool and the MT pool are held constant according to literature or pre-acquisition measurements.

Note the amplitude of RF pulse is incorporated into the lineshape function. Consequently, to calculate $T_{1D}$ from $RATIO_{dosl}$, a $B_1$ map is acquired to convert the designed spin-lock RF amplitude into the actual applied RF amplitude. Given the measured $RATIO_{dosl}^{acq}$ and the corresponding local $B_1$ value, the optimal entry in the dictionary is identified by minimizing the residual between $RATIO_{dosl}^{acq}$ and the simulated $RATID_{dosl}$. The $T_{1D}$ value associated with this best match is then assigned as the estimated $T_{1D}$.

# 3.3 Simulation studies

The simulation studies were conducted to validate the proposed theory and the acquisition approach. All simulations were performed using custom MATLAB code, which are available on GitHub (https://github.com/zjgao-spin/Dipolar-order-mapping).

## 3.3.1 Simulation study 1: Accuracy of quantification

### 3.3.1.1 Accuracy of approximated $RADIO_{dosl}$

To assess the accuracy of this analytical expression of $RATIO_{\mathrm{dosl}}$ at Eq. 10, we compared it against the exact numerical solution of $RATIO_{dosl}$ obtained by integrating the Bloch-McConnell-Provotorov equations. For this comparison, we utilized tissue parameters for white matter derived from previous publications[33,34]: $T_{1a} = 1840ms$, $T_{1b} = 340ms$, $T_{2a} = 69ms$, $T_{2b} = 10\mu s$, $MPF = 15\%$, and $R = 23s^{-1}$. We evaluated $T_{1D}$ values of 2, 4, 6, and 8 ms, consistent with the previously reported range for white matter[6,14]. The ranges for $\Delta\omega^{d(1)}$ and $\omega_1^{d(1)}$ were set to cover common experimental values: 2-7 kHz and 200-800 Hz, respectively. Furthermore, to theoretically validate the sensitivity of the method, we analyzed the relationship between $RATIO_{dosl}$ and $T_{1D}$ across a range of 1 to 10 ms. Additionally, we investigated how $RATIO_{\mathrm{dosl}}$ varies with other tissue parameters, including $R_{1a}$ (0.3–1.5 Hz), $R_{2a}$ (10–70 Hz), $R_{1b}$ (2–5 Hz), $R$ (15–25 Hz), $MPF$ (10–20%), and $T_{2b}$ (9–11 μs).

### 3.3.1.2 Accuracy of $T_{1D}$ quantification

The analytical estimation does not require generating a dictionary. However, when using this method, the approximate $RATIO_{\mathrm{dosl}}$ model (Eq. 10) can introduce bias in the estimated $T_{1D}$ due to underlying theoretical assumptions. This bias between the estimated and ground-truth $T_{1D}$ can be mitigated by the appropriate selection of acquisition parameters.

To identify parameter regimes that minimize the bias when using the approximate model, we computed the error in the estimated $T_{1D}$ relative to a ground-truth value of $T_{1D}$= 6.2 ms[6]. Acquired $RATIO_{\text{dosl}}$ values were generated via numerical simulations using Eqs. 1–5 and 11. The switch times, $T_s$, for the dual and single-frequency spin-lock pulses were chosen as 0.5 ms and 40 ms, respectively, given that white matter $T_{1D}$ typically ranges from 3 to10 ms[7]. The simulation parameters were set as follows: $\Delta\omega^{d(1)}/2\pi$ from 2 to 7 kHz, $\omega_1^{d(1)}/2\pi$ ranged from 200 to 800 Hz, TSL from 40 to 120 ms., and $\frac{\Delta\omega^{d(1)}}{\Delta\omega^{d(2)}} = \frac{\omega_1^{d(1)}}{\omega_1^{d(2)}} = 5$. The $T_{1D}$ values were then calculated by solving Eq.10.

## 3.3.2 Simulation study 2: Robustness of the proposed method

### 3.3.2.1 Robustness in presence of B1 and B0 inhomogeneity

$B_1$ and $B_0$ field inhomogeneities are common in MRI systems. We performed simulations to investigate the robustness of the proposed method in the presence of these inhomogeneities. The simulations utilized the same spin-lock pulse and tissue parameters as in Simulation Study 1, with $\Delta\omega^{d(1)}/2\pi$, $\omega_1^{d(1)}/2\pi$ and TSL fixed at 5 kHz, 500 Hz and 80 ms, respectively. $B_1$ and $B_0$ were varied over ranges of 0.7-1.3 n.u. and -100 to 100 Hz, respectively.

Furthermore, we compared the mean and standard deviation of the estimated $T_{1D}$ with and without $B_1$ inhomogeneity correction. In this analysis, $B_1$ and $B_0$ values were sampled 10,000 times from a uniform distribution over respective range. $T_{1D}$ was estimated using both analytical estimation and dictionary matching. The dictionary was generated covering a $B_1$ range of 0.7 to 1.3 (step size $1 \times 10^{-4}$) and a $T_{1D}$ range of 0 to 15 ms (step size $1 \times 10^{-5}$ms).

### 3.3.2.2 Robustness against variations of the tissue parameters

Like other quantitative MT methods, we assumed that certain MT parameters remain constant when calculating $T_{1D}$. $RATIO_{\text{dosl}}$ is predicted to have low sensitivity to most MT parameters, with the exception of $T_{2b}$. In this study, we performed simulations to investigate the errors in the estimation of $T_{1D}$ when MT parameters, specifically $R_{1b}$, $R$, and MPF, are not constant. We conducted 10,000 simulations with parameters randomly sampled within the following ranges: $R_{1b} = 2$–5 Hz, $R$ = 15–25 Hz, and MPF = 10–20%. We then calculated the variation of the estimated $T_{1D}$ under these parameter variations.

#### 3.3.2.3 Robustness against noise

Quantitative imaging requires a sufficient signal-to-noise ratio (SNR). In this study, we investigated the influence of SNR on $RATIO_{\mathrm{dosl}}$ and $T_{1D}$ estimation. Simulated MRI signals $M_d^{(1)}$, $M_d^{(2)}$, and $M_s^{(1)}$ were corrupted with additive white Gaussian noise at SNR levels ranging from 20 to 100. For each SNR level, 10,000 random experiments were performed, and the resulting distributions of the estimated $T_{1D}$ were analyzed.

## 3.4 Phantom and in-vivo studies

### 3.4.1 Preparation of phantoms and health volunteer.

Agar phantoms and Prolipid 161 (PL161; Ashland Specialty Ingredients, USA) phantoms were prepared for this study and underwent the same MRI protocol. The agar phantoms were prepared at 1%, 2%, 3%, and 4% (w/w), while the PL161 phantoms were mixed with pure water at 4%, 8%, 12%, and 16% (w/w). PL161 exhibits strong ihMT contrast and was therefore regarded as a validation of the ihMT effect[5,20].

Ten healthy volunteers (age range 25-30 years; 5 male and 5 female) were enrolled in this study under the approval of our Institutional Review Board (Ref No. 2016.150). Exclusion criteria included a history of neurological diseases, brain injury, major psychiatric illness, or drug or alcohol misuse. The study was performed in accordance with the institutional ethical guidelines and the ethical standards of the 1964 Declaration of Helsinki and its subsequent amendments. Written informed consent was obtained from all participants. Each volunteer underwent test-retest MRI examinations with a 7-10 day interval.

### 3.4.2 MRI protocol

All MRI data acquisitions were performed using a 3T Prisma scanner (Siemens Healthineers, Germany) equipped with a 64-channel head-neck receiver coil at the room temperature (~20°C). The MRI scan protocol for in-vivo study included the following parameters with the identical FOV 260 × 260 mm$^2$ :

(1) A 3D $T_1$-weighted axial image was acquired for anatomical imaging using magnetization prepared rapid gradient echo (MP-RAGE) sequence with the following parameters: TE = 1.67 ms, TR = 1900 ms, voxel size = 1.5 × 1.5 × 2.5 mm$^3$, and a scan time of 2 minutes, 3 seconds.

(2) A 3D $B_1$ map was obtained using the Siemens clinical brain protocol. For $B_1$ mapping, the voxel size was 2.9 × 2.9 × 2.5 $mm^3$ with a 57 s acquisition.

(3) A diffusion tensor imaging (DTI) scan was performed with TE = 77 ms, TR = 3200 ms, voxel size = 2.5 × 2.5 × 2.5 $mm^3$, b-value = 0 $s/mm^2$ and 1000 $s/mm^2$, 30 diffusion directions, and a scan time of 3 minutes, 58 seconds.

(4) MPF and $T_{1D}$ are acquired in a single scan. Spin-lock prepared images were acquired based on 2D turbo spin echo (TSE) sequence. The spin-lock parameters were set as follow: $\Delta\omega^{d(1)}/2\pi$=5kHz , $\omega_1^{\mathrm{d}(1)}/2\pi$=500Hz , $\Delta\omega^{d(1)} = \Delta\omega^{s(1)}$, $\omega_1^{\mathrm{d}(1)} = \omega_1^{\mathrm{s}(1)}$, $\frac{\Delta\omega^{d(1)}}{\Delta\omega^{d(2)}} = \frac{\omega_1^{d(1)}}{\omega_1^{d(2)}} = 5$, TSL = 80ms. Switch times, $T_s$ , are 0.5 ms for dual-frequency spin-lock and 40 ms for single-frequency spin-lock. The voxel size was 1.5 × 1.5 × 5 $mm^3$, the number of slices was 3, TE was 9.2 ms, and TR was 2200 ms. Magnetization reset with $T_1$ recovery time was 1800 ms for fast acquisition strategy[28]. The acquisition time was 1 min 2s for 3 spin-lock prepared images per slice.

In addition, one volunteer underwent Z-spectroscopic data acquisition using an MT-weighted spoiled gradient echo (GRE) sequence with a Gaussian pulse for off-resonance saturation with 11 Δ values (2, 3, 4, 6, 8, 12, 16, 20, 32, and 36 kHz) and an independence $R_1$ maps acquisition to calculate the MT parameters[27,35].

Notably, the MPF and $T_{1D}$ acquisitions in phantom study were performed using the same parameter as in (4), except a FOV of 240 mm×240mm, a voxel size of 2×2×5 $mm^3$.

### 3.4.3 Data processing and analysis

To convert $RATIO_{\mathrm{dosl}}$ to $T_{1D}$, we assumed that the MT model parameters (MPF, $R_{1b}$, $T_{2b}$, and $R$) remained constant. For agar phantoms, we used MPF = 2%, $R_{1b} = 1\,Hz$, $T_{2b} = 10\,\mu s$ and $R = 210 s^{-1}$ [36]. For PL161 phantoms, we used MPF = 15%, $R_{1b} = 5\,Hz$, $T_{2b} = 17\,\mu s$ and $R = 46 s^{-1}$, respectively[19]. For healthy volunteers, we used the qMRLab (https://qmrlab.org/) to fit the Z-spectroscopic data from one volunteer and estimated MPF=13.6%. $R = 20 s^{-1}, T_{2b} = 9.7 \mu s$ , while $R_{1b}$ was fixed at $2.9 Hz$ based on literature[34].

$RATIO_{\mathrm{dosl}}$ maps were calculated from three spin-lock prepared images $M_d^{(1)}$, $M_d^{(2)}$ and $M_s^{(1)}$ via Eq. 11. Prior to this calculation, all spin-lock prepared images were smoothed using a mean filter with a kernel size of 4 to reduce noise. $T_{1D}$ maps were then obtained from $RATIO_{\mathrm{dosl}}$ using both analytical estimation and dictionary matching method for the *in vivo* study, respectively. The dictionary was generated over a $T_{1D}$ range of 0 to 30 ms (step size $1 \times 10^{-5}$ms) and a $B_1$ range of 0.7 to 1.3 (step size $1 \times 10^{-4}$). Note that only dictionary matching was applied for the phantom studies, as the accuracy of analytical estimation for the phantom under the current spin-lock parameters requires further investigation.

$R_{\mathrm{mpfsl}}$ maps were derived from $M_d^{(1)}$ and $M_d^{(2)}$, and converted to MPF maps using the fast MPF-SL method with a dictionary approach[32,37]. The in-plane resolution of the $B_1$ map and DTI data was resampled to 1.5 × 1.5 mm$^2$ using linear interpolation to match the $T_{1D}$ and MPF data. To analyze $T_{1D}$ maps in ROIs of white matter, the $T_1$-weighted images and DTI data were used for fiber bundles segmentation. The TractSeg opensource tool (https://github.com/MIC-DKFZ/TractSeg) was employed to segment the fiber bundles of white matter [37]. In this study, the acquired slices for MPF measurement primarily included 16 regions of white matter fiber bundles: Arcuate fascicle (AF_left, AF_right), Anterior Thalamic Radiation (ATR_left, ATR_right), Corpus Callosum Genu (CC_2), Corpus Callosum Rostral body (CC_3), Corpus Callosum Posterior midbody (CC_5), Corpus Callosum Splenium (CC_7), Cingulum (CG_left, CG_right), Optic radiation (OR_left, OR_right), Middle longitudinal fascicle (MLF_left, MLF_right), and Fronto-pontine tract (FPT_left, FPT_right). The mean and standard deviation of $T_{1D}$, as well as MPF and $RATIO_{\mathrm{dosl}}$, were calculated for each white matter bundles. Additionally, the assessment of test-retest reproducibility of the *in vivo* study was performed, as described in Supporting Information 1.

# 4.Results

## 4.1 Simulation studies

Figure 3 compares the approximated analytical $RATIO_{\mathrm{dosl}}$ with its exact numerical solution. As shown in Figures 3(a) and (b), the approximation (markers) demonstrates excellent agreement with the exact solution (solid lines). Under appropriate spin-lock pulse parameters, the relative error remains below 1% (Figure 3(c)). These results confirm that the proposed approximation provides a reliable estimate, supporting its utility in practical applications.

Figure 4(a) illustrates the sensitivity of $RATIO_{\mathrm{dosl}}$ to $T_{1D}$. With $\omega_1^{d(1)}/2\pi$ fixed at 500 Hz and $\Delta\omega^{d(1)}/2\pi$ set to 5, 6, and 7 kHz, $RATIO_{\text{dosl}}$ increases markedly as $T_{1D}$ rises from 1 to 10 ms, demonstrating high sensitivity. Figure 4(b) depicts the sensitivity of $RATIO_{\text{dosl}}$ to other tissue parameters including $R_{1a}$, $R_{2a}$, $R_{1b}$, MPF, $R$, and $T_{2b}$. $RATIO_{\text{dosl}}$ is essentially independent of $R_{1a}$, $R_{2a}$, $R_{1b}$, and MPF. While it exhibits low sensitivity to the exchange rate $R$ and pronounced sensitivity to $T_{2b}$, these parameters are typically treated as constants in human studies.

Figure 5 presents an accuracy analysis of $T_{1D}$ estimation from measured $RATIO_{\text{dosl}}$ using analytical estimation method across various spin-lock parameters ($\Delta\omega^{d(1)}$, $\omega_1^{d(1)}$, and TSL). Figures 5(a) and 5(d) suggest that $\Delta\omega^{d(1)}/2\pi$ values of 3-7 kHz combined with $\omega_1^{d(1)}/2\pi$ = 500 Hz are optimal, yielding relative errors below 5%. Figures 5(b) and 5(e) indicate that $\omega_1^{\text{d}(1)}/2\pi$ values of 500-700 Hz are favorable. The relationship between relative error and TSL varies by $\Delta\omega^{d(1)}$; notably, at $\Delta\omega^{d(1)}/2\pi$ = 5 kHz, the error is nearly independent of TSL. Based on these findings, we selected $\omega_1^{d(1)}/2\pi$ = 500 Hz, $\Delta\omega^{d(1)}/2\pi$ = 5 kHz, and TSL = 80 ms for *in vivo* experiments. These parameters fall within the scanner's SAR constraints and RF hardware limits.

Figure 6(a) displays the dependence of $RATIO_{\text{dosl}}$ on $B_0$ and $B_1$ inhomogeneity, alongside the relative error compared to the ground truth ($B_0$= 0 Hz, $B_1$= 1 n.u.). For $B_0$ between –100 and 100 Hz, $RATIO_{\text{dosl}}$ exhibits minimal variation and minor oscillations around the ground truth. However, consistent with theoretical predictions, $RATIO_{\text{dosl}}$ is sensitive to $B_1$ inhomogeneity because the lineshape depends on the actual RF amplitude. Notably, incorporating $B_1$ maps for retrospective correction of $T_{1D}$ effectively mitigates the impact of this inhomogeneity. As shown in Table.1A, the bias in $T_{1D}$ quantification due to field inhomogeneity is significantly reduced following this correction. Table.1B details the results under variations of MT parameters. The distributions of $RATIO_{\text{dosl}}$, $T_{1D}$ (analytical estimation), and $T_{1D}$ (dictionary matching) are centered at $0.41 \pm 0.02$, $6.29 \pm 0.34$ ms, and $6.21 \pm 0.40$ ms, respectively. Estimated $T_{1D}$ values show minimal bias (<0.1 ms) relative to the ground truth (6.2 ms). These findings indicate that, assuming fixed MT parameters (excluding $T_{2b}$), the estimation of $T_{1D}$ is insensitive to variations in MT parameters.

Figure 6(b) presents the robustness against SNR. We calculated the median due to the presence of outliers at low SNR and relative bias for the estimated $T_{1D}$. This analysis confirms that SNR is critical for reliability, indicating that analytical estimation and dictionary matching methods require an SNR of at least ~40 to maintain a relative bias below 3%.

## 4.2 Phantom and in-vivo studies

Figure 7(a) shows the MPF-SL acquisitions for the agar and PL161 phantoms. The $R_{\mathrm{mpfsl}}$ and the derived MPF are strongly associated with phantom concentration. Figure 7(b) presents the corresponding $RATIO_{\mathrm{dosl}}$ and $T_{1D}$ maps. The $RATIO_{\mathrm{dosl}}$ map highlights the contrast of the PL161 phantom, demonstrating its sensitivity to the ihMT effect. In the $T_{1D}$ maps, the long $T_{1D}$ of the PL161 phantom is confirmed by our method, whereas the agar phantom exhibits a notable MPF but negligible $T_{1D}$, consistent with its lack of dipolar behavior. Figures 7(c–f) show the relationships of $RATIO_{\mathrm{dosl}}$ and $T_{1D}$ with phantom concentration. Both $RATIO_{\mathrm{dosl}}$ and $T_{1D}$ show no obvious dependence on PL161 concentration overall.

Figure 8 presents *in vivo* results from one volunteer (V1). The $T_1$-weighted anatomical image for the acquired slices and the 16 major white matter bundles are shown in Figures 8(a) and 8(b). Figures 8(c) and 8(d) show the MPF maps and $RATIO_{dosl}$ maps, which exhibit different contrasts for white matter and indicate that these two parameters may carry different molecular signatures of tissues. $T_{1D}$ maps were derived from the $RATIO_{dosl}$ maps using analytical estimation and dictionary matching, with and without $B_1$ correction. The resulting $T_{1D}$ maps preserve a contrast similar to that of the $RATIO_{dosl}$ maps (Fig. 8(e–h)). Results for the other volunteers are provided in Supporting Information 2. Table 2 summarizes the mean and standard deviation of MPF, $RATIO_{dosl}$, and the corresponding $T_{1D}$ across the 16 major white matter fiber bundles in 10 volunteers.

# 5.Discussion

Our proposed framework enables simultaneous quantification of $T_{1D}$ and MPF within a single scan. While MPF primarily reflects macromolecular content, $T_{1D}$ offers complementary sensitivity to microstructural integrity via the ihMT effect. Central to this approach is $RATIO_{\mathrm{dosl}}$, a variable specific to $T_{1D}$, calculated from the distinct relaxation rate $R_{dosl}$. Deriving $RATIO_{\mathrm{dosl}}$, and subsequently converting it to $T_{1D}$ with $B_1$ correction, requires only three spin-lock prepared images. Notably, two of these images can be used for MPF quantification without additional data acquisition[28]. Consequently, this simultaneous mapping provides a comprehensive characterization of both tissue microstructure and macromolecular content.

Regarding previous literature, Varma et al. reported a $T_{1D}$ value of approximately 6.2 ms in *in vivo* white matter using a multi-ihMTR approach[6], whereas West et al. reported $T_{1D}$ values of approximately 3.5–5.5 ms in white matter using an MRF technique[20]. In this study, we obtained white matter $T_{1D}$ values in the range of ~3.70–4.80 ms based on dictionary matching with $B_1$ correction. Although these values are broadly consistent with prior reports, precise validation of the true $T_{1D}$ remains challenging. We additionally performed *in vivo* experiments varying the $T_s$ of dual frequency spin-lock from 0.5 to 20 ms. Results indicated that white matter was strongly highlighted when the dual-frequency spin-lock $T_s$ was ≤ 4 ms (Figure S1.2 in Supporting Information 1), providing indirect validation for our obtained $T_{1D}$ values based on $T_{1D}$ filtering.

For the proposed method, because $RATIO_{\mathrm{dosl}}$ is derived from different $R_{1\rho}$ measurements under a constant spin-lock field direction, contributions from the water pool are effectively cancelled. Simulation studies further indicate that $RATIO_{\mathrm{dosl}}$ exhibits only weak dependence on other MT parameters (e.g., MPF, R, and $R_{1b}$).

Prolonged scan times often limit the clinical utility of $T_{1D}$ assessment[6,20]. Our approach addresses this by requiring only three spin-lock prepared images for joint $T_{1D}$ and MPF quantification. Although this study demonstrated the technique in 2D slices, extending the method to whole-brain coverage is feasible using fast 3D MRI acquisition strategies[38–40]. In previous work[40], using four spin-lock prepared images combined with 3D FSE/TSE acquisition, whole-brain coverage of $T_{1\rho}$ quantification could be achieved within 5 minutes. Consequently, by leveraging strategies that rely solely on three spin-lock prepared images, we estimate the proposed method can achieve whole-brain coverage in 5 minutes.

Our method employs the least negative eigenvalue to model signal evolution, an approach traditionally assumed to necessitate long saturation times[30]. However, our analysis indicates that this approximation remains robust at short spin-lock durations (e.g., 80 ms) under specific experimental conditions. By utilizing large frequency offsets relative to the spin-lock amplitude ($\Delta\omega \gg \omega_1$), the system enters a regime where magnetization decay is dominated by a single component. In this context, contributions from faster-decaying modes are negligible, allowing the least negative eigenvalue to accurately characterize signal evolution as a mono-exponential process.

In this study, we compared analytical estimation and dictionary matching for $T_{1D}$ quantification. At SNR $\geq 40$, both approaches yielded low error; analytical estimation showed a small residual bias due to model approximations, whereas dictionary matching achieved slightly higher accuracy. Analytical estimation offers fast, continuous estimates without the memory and runtime overhead associated with large dictionaries; however, its accuracy can depend on careful optimization of acquisition parameters to control bias across tissue types. Dictionary matching is, in principle, applicable to arbitrary acquisition settings. Overall, the choice of $T_{1D}$ estimation approach should be guided by the study's SNR conditions and acquisition design.

Looking ahead, other spin-lock-based quantitative MT techniques, such as pulsed spin-lock approaches[25], may also be leveraged for dipolar-order quantification. Pulsed spin-lock methods can mitigate RF hardware limitations, offering particular benefits for body imaging and low-field applications where power constraints are pronounced.

Although our theoretical analysis and experimental results support the reliability and clinical feasibility of spin-lock–based $T_{1D}$ quantification, several areas warrant further investigation to refine the technique:

**Modeling Assumptions:** First, we treated the lineshape parameter $T_{2b}$ as a constant. While this is commonly used in MT modeling, our simulations indicate that $RATIO_{\mathrm{dosl}}$ is sensitive to this parameter, suggesting that the impact of $T_{2b}$ variations on $T_{1D}$ accuracy deserves closer examination. Furthermore, our current model assumes a single $T_{1D}$ component and a dipolar order fraction of unity. Since previous studies suggest that multi-component models and variable dipolar fractions influence the ihMT effect[41,42], future work should explore more comprehensive models to fully characterize the underlying physical mechanisms.

**Biological and Physiological Factors:** The potential orientation dependence of this specific spin-lock approach remains to be determined, given conflicting evidence in prior studies[26,27,43]. Additionally, physiological states such as metabolic changes or altered arousal could theoretically modulate water-myelin interactions, representing another avenue for investigation.

**Optimization and Validation:** While our acquisition parameters were chosen to satisfy analytical approximations and yielded reliable results, further optimization could simultaneously maximize $T_{1D}$ sensitivity and SNR within hardware constraints. Finally, establishing a standardized benchmark is crucial. Future studies should validate the proposed method against existing $T_{1D}$ quantification techniques and correlate findings with histology and clinical pathologies (e.g., demyelination) to firmly establish its utility in routine practice.

# 6.Conclusion

We successfully demonstrated a novel and rapid off-resonance spin-lock technique for quantifying the dipolar relaxation time $T_{1D}$. The capacity to measure both MPF and $T_{1D}$ in a single scan highlights the clinical utility of this approach, offering a promising pathway for integrating molecular microstructure imaging into clinical diagnostics.

# Acknowledgment

This study was supported by a grant from the Research Grants Council of the Hong Kong SAR (Project GRF 14213322) and a grant from the Innovation and Technology Commission of the Hong Kong SAR (Project No. MRP/046/20x).

# Conflict of interest statement

Weitian Chen is a shareholder of Illuminatio Medical Technology Limited.

# Figure and Table

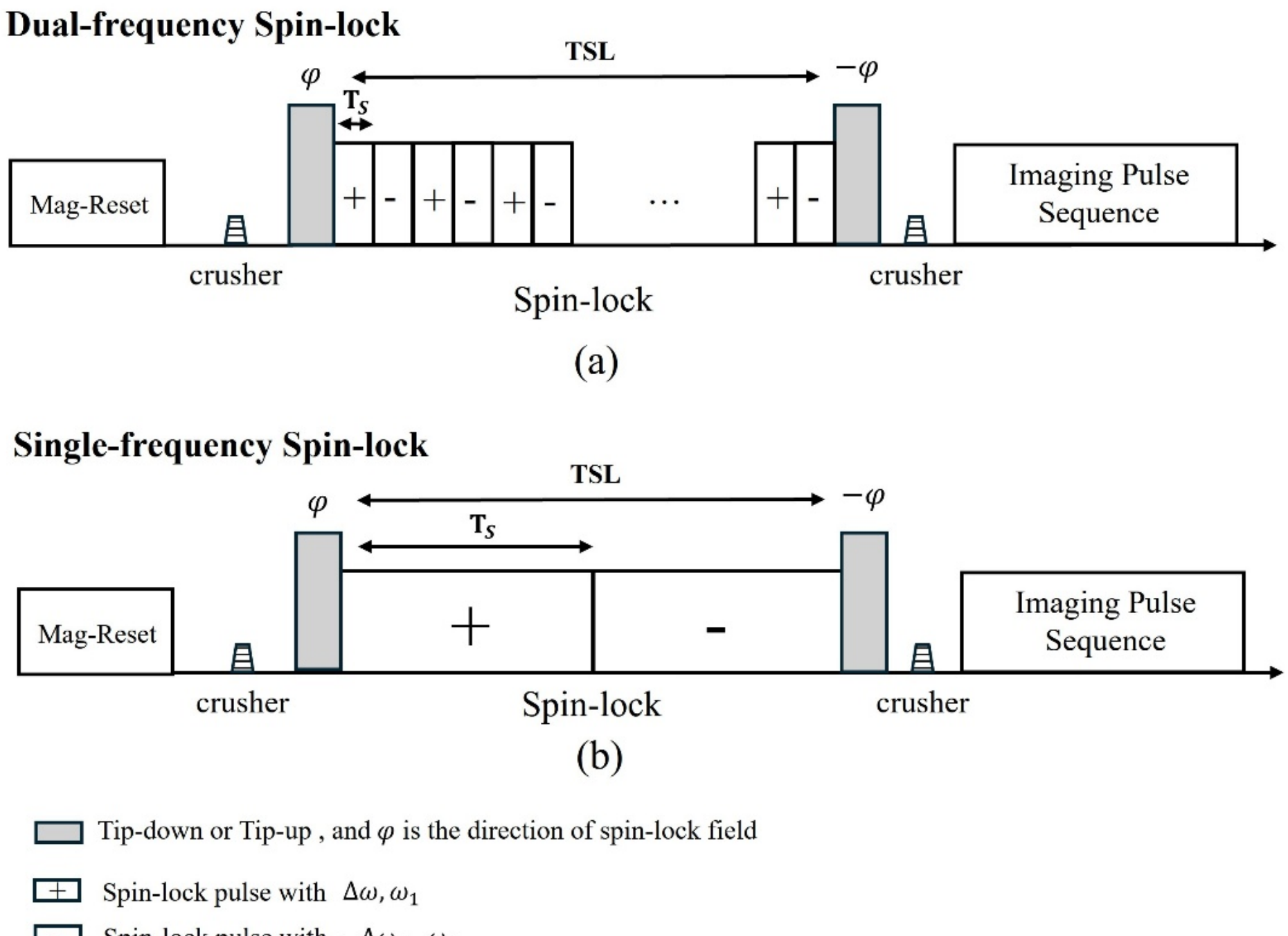

**Figure 1.** (a) Dual-frequency spin-lock pulse sequence with a short switch time $T_s$ (shorter than the tissue $T_{1D}$). (b) Single-frequency spin-lock pulse sequence with a long switch time $T_s$ (longer than the tissue $T_{1D}$).

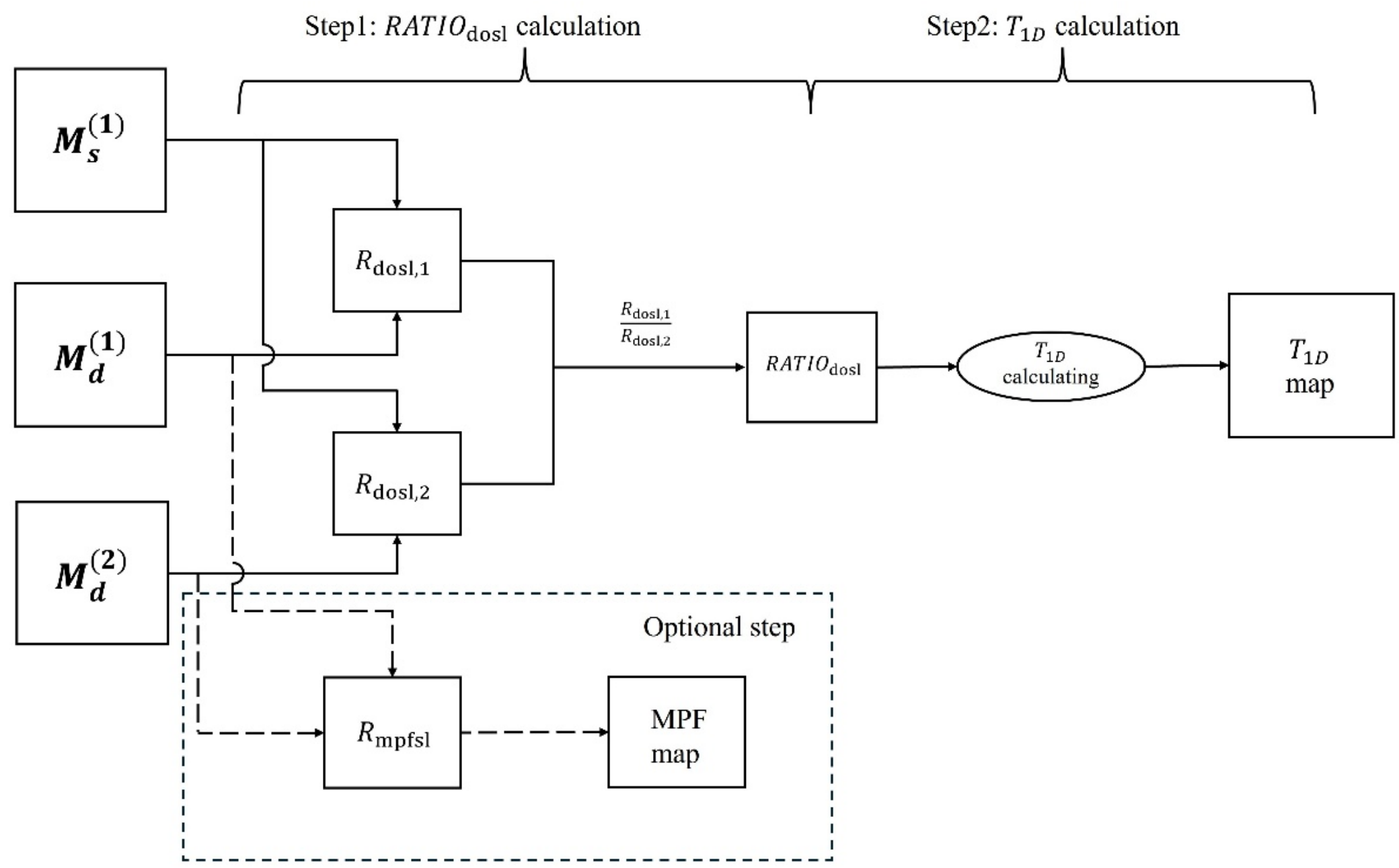

**Figure 2.** Workflow of the $T_{1D}$ calculation. In Step 1, $RATIO_{\mathrm{dosl}}$ is calculated from three spin-lock prepared images, $M_s^{(1)}$, $M_d^{(1)}$, and $M_d^{(2)}$. In Step 2, $T_{1D}$ is derived from the $RATIO_{\mathrm{dosl}}$ values. An MPF map can be obtained from $M_d^{(1)}$and $M_d^{(2)}$ in an optional step.

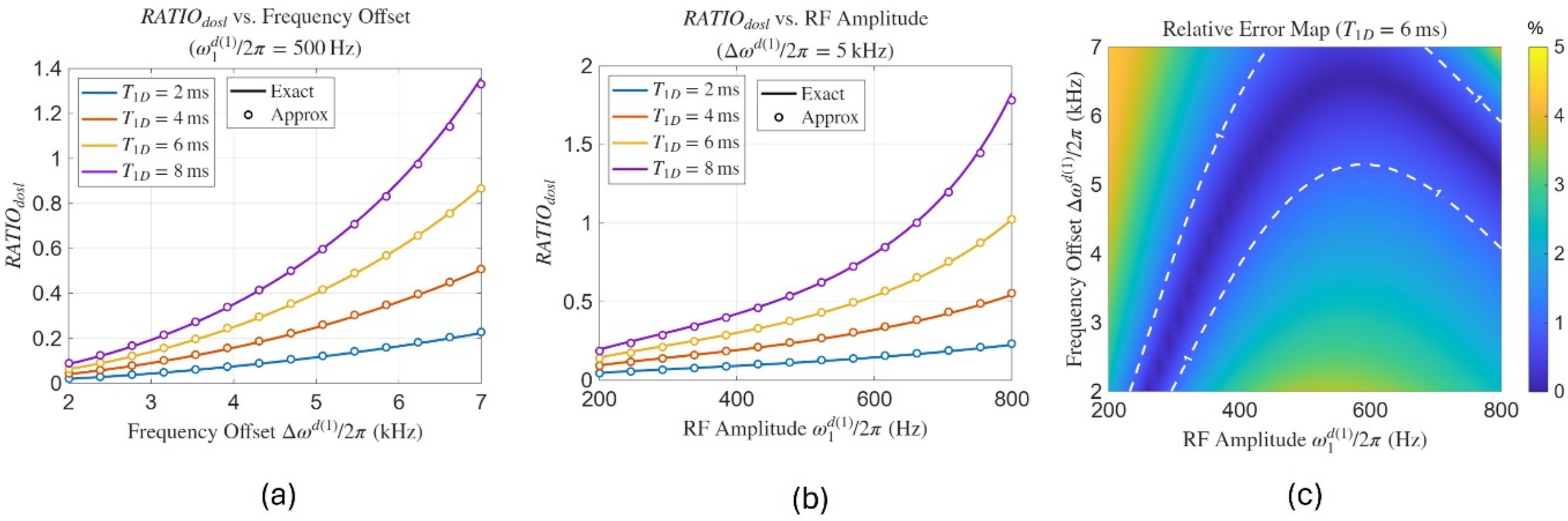

**Figure 3.** Comparison between the approximated analytical $RATIO_{dosl}$ and its exact numerical solution. (a) The relationship between $RATIO_{dosl}$ and $\Delta\omega^{d(1)}/2\pi$ (2-7kHz) at a fixed $\omega_1^{\mathrm{d}(1)}/2\pi$ of 500 Hz. For each $T_{1D}$ (2, 4, 6, 8 ms), the approximate results (markers) closely match the numerical solution curves (solid lines). (b) The relationship between $RATIO_{dosl}$ and $\omega_1^{\mathrm{d}(1)}/2\pi$ (200–800 Hz) at a fixed $\Delta\omega^{d(1)}/2\pi$ of 5 kHz. The same agreement between approximate (marker) and exact numerical (solid lines) results is observed. (c) The relative error across the $\omega_1^{\mathrm{d}(1)}/2\pi$ and $\Delta\omega^{d(1)}/2\pi$ range.

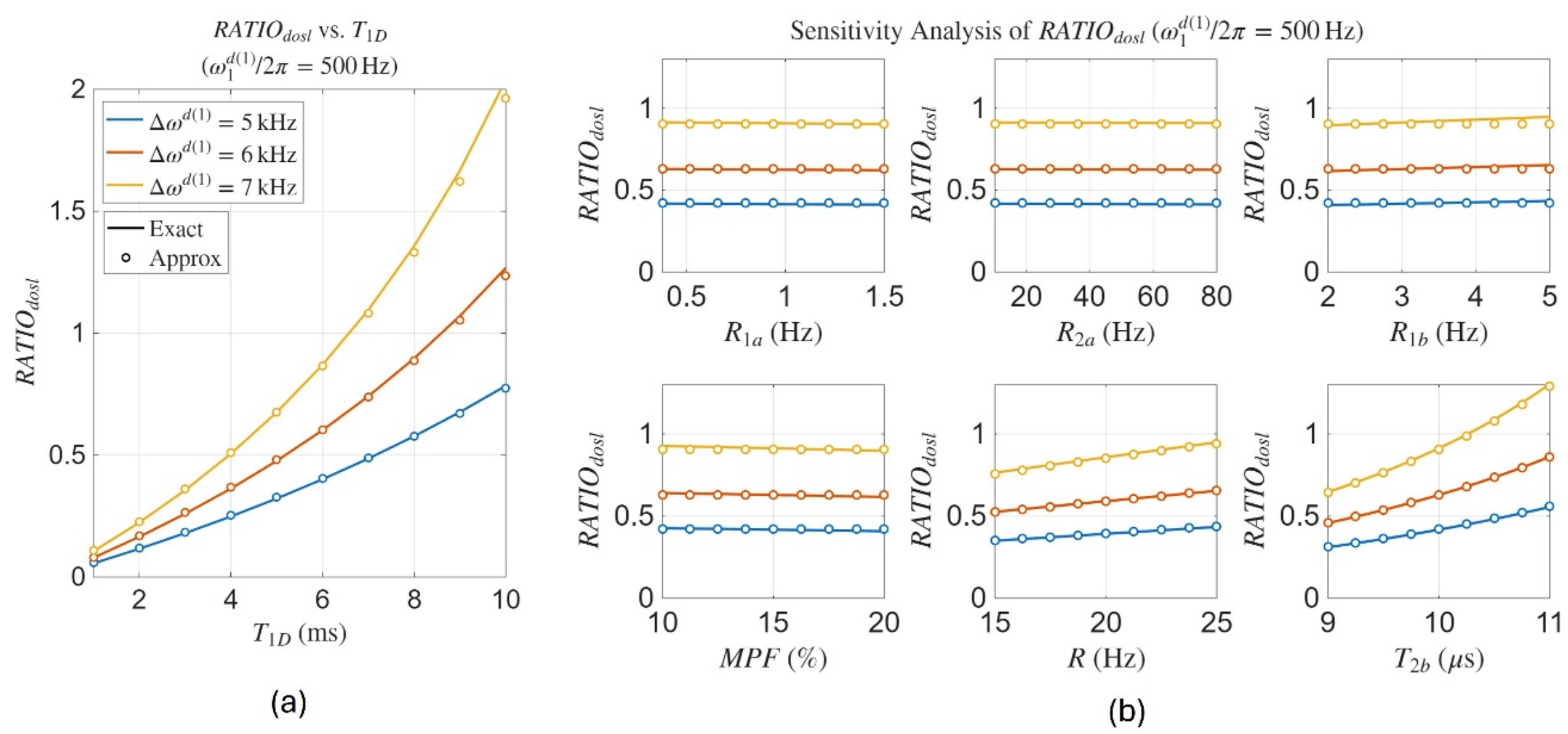

**Figure 4.** (a) The relationship between $RATIO_{dosl}$ and $T_{1D}$ at fixed $\omega_1^{d(1)}/2\pi$ = 500Hz with selected $\Delta\omega^{d(1)}/2\pi$ = 5,6, and 7 kHz. Approximate results (markers) closely match numerical solutions (solid lines) over $T_{1D}$ = 1–10 ms. For all $\Delta\omega^{d(1)}$ values, $RATIO_{dosl}$ increases with $T_{1D}$, confirming the high sensitivity of $RATIO_{dosl}$ to $T_{1D}$. (b) The relationship between $RATIO_{dosl}$ and other tissue parameters, including $R_{1a}$, $R_{2a}$, $R_{1b}$, MPF, $R$, and $T_{2b}$.

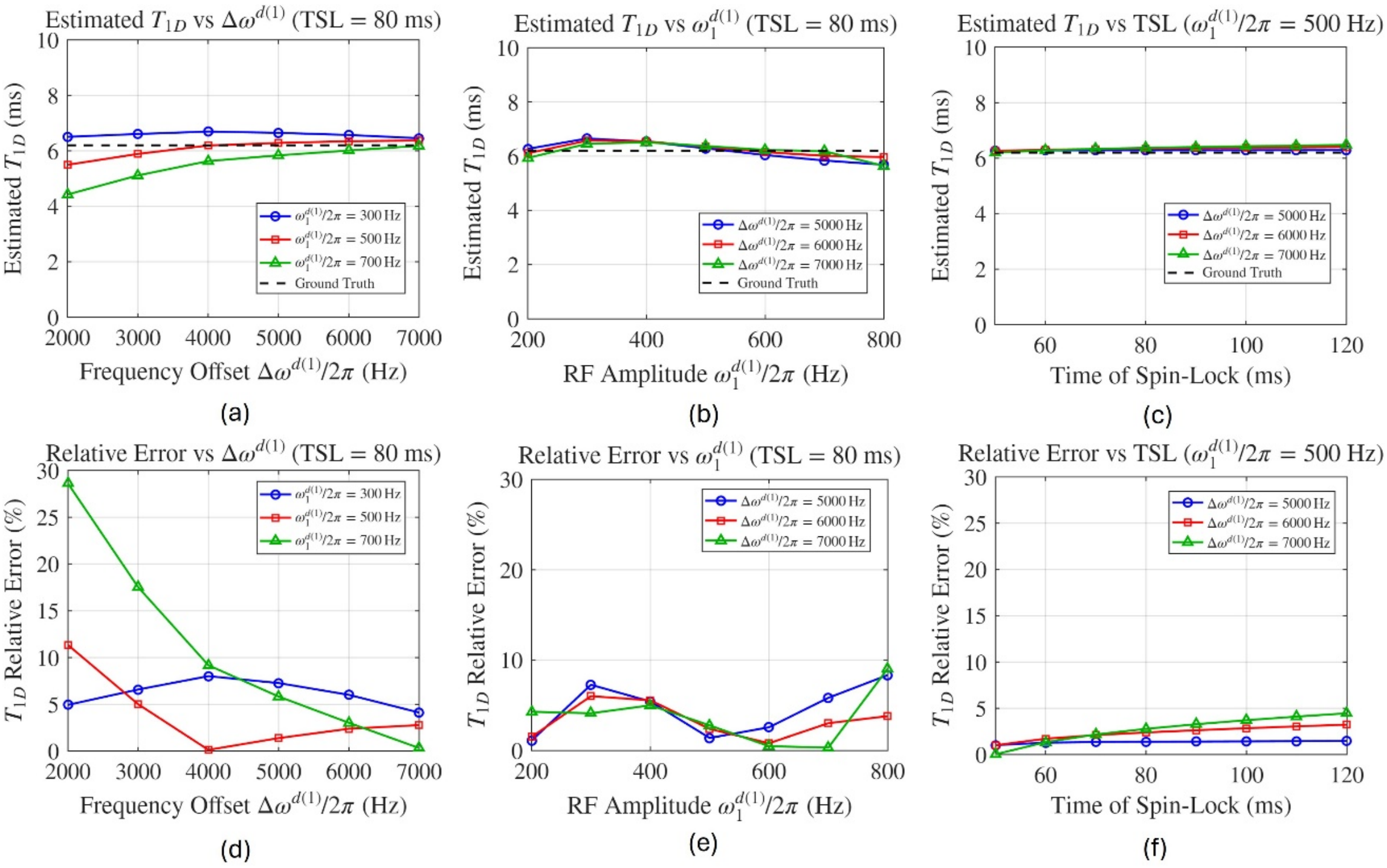

**Figure 5.** Simulation of the accuracy of $T_{1D}$ estimation as a function of spin-lock pulse parameters. (a) Estimated $T_{1D}$versus resonance frequency offset ($\Delta\omega^{d(1)}/2\pi$, 2–7 kHz) at fixed TSL = 80 ms and selected spin-lock field strength $\omega_1^{d(1)}/2\pi$ values of 300, 500, and 700 Hz. (b) Estimated $T_{1D}$versus spin-lock field strength ($\omega_1^{d(1)}/2\pi$, 200–800 Hz) at fixed TSL = 80 ms and selected $\Delta\omega^{d(1)}/2\pi$ values of 5, 6, and 7 kHz. (c) Estimated $T_{1D}$ versus spin-lock duration (TSL, 40–120 ms) at fixed $\omega_1^{d(1)}/2\pi$ = 500 Hz and selected ($\Delta\omega^{d(1)}/2\pi$ values of 5, 6, and 7 kHz. (d-f) Corresponding relative errors of the estimated $T_{1D}$with respect to the ground-truth value. The dashed line in (a–c) indicates the ground-truth $T_{1D} = 6.2$ms.

**Robustness against $B_0$ and $B_1$ inhomogeneity**

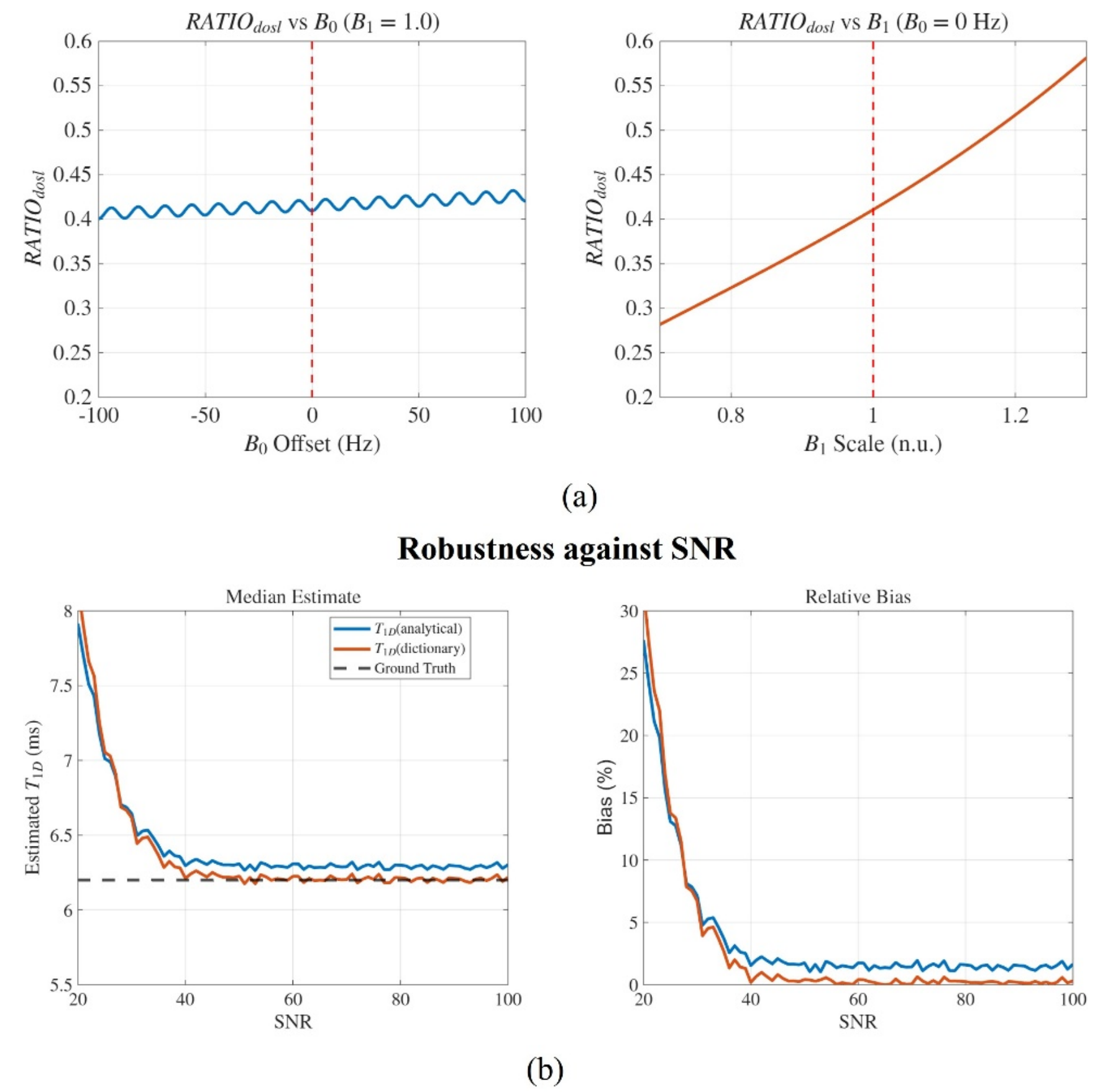

**Figure 6.** (a) Robustness of $RATIO_{dosl}$ to $B_0$ and $B_1$ inhomogeneity, displaying sensitivity to $B_0$ offsets ($-100$ to 100 Hz) and $B_1$ scaling factors (0.7 to 1.3 n.u.). (b) Robustness to SNR: a comparison of analytical estimation versus dictionary matching for $T_{1D}$. The plot shows the median estimate and relative bias compared to the ground-truth $T_{1D}$= 6.2ms across an SNR range of 20–100.

**Table1. Summary of robustness tests for $T_{1D}$ estimation.**

A. Robustness against $B_1$ and $B_0$ inhomogeneity

| $B_1$ (n.u.) | $B_0$ (Hz) | $RATIO_{dosl}$ | No $B_1$ correction | | $B_1$ correction | |
|---|---|---|---|---|---|---|
| | | | $T_{1D}$ (ms) (Analytical estimation) | $T_{1D}$ (ms) (Dictionary matching) | $T_{1D}$ (ms) (Analytical estimation) | $T_{1D}$ (ms) (Dictionary matching) |
| 0.7-1.3 | -100-100 | 0.42±0.09 | 6.52±1.31 | 6.53±1.56 | 6.38±0.20 | 6.24±0.26 |

B. Robustness against MT parameters (excluding $T_{2b}$)

| $R_{1b}$ (Hz) | MPF (%) | R (Hz) | $RATIO_{dosl}$ | $T_{1D}$ (ms) (Analytical estimation) | $T_{1D}$ (ms) (Dictionary matching) |
|---|---|---|---|---|---|
| 2-5 | 10-20% | 15-30 | 0.41±0.02 | 6.29±0.34 | 6.21±0.40 |

Note: the ground truth $T_{1D} = 6.2\ ms$

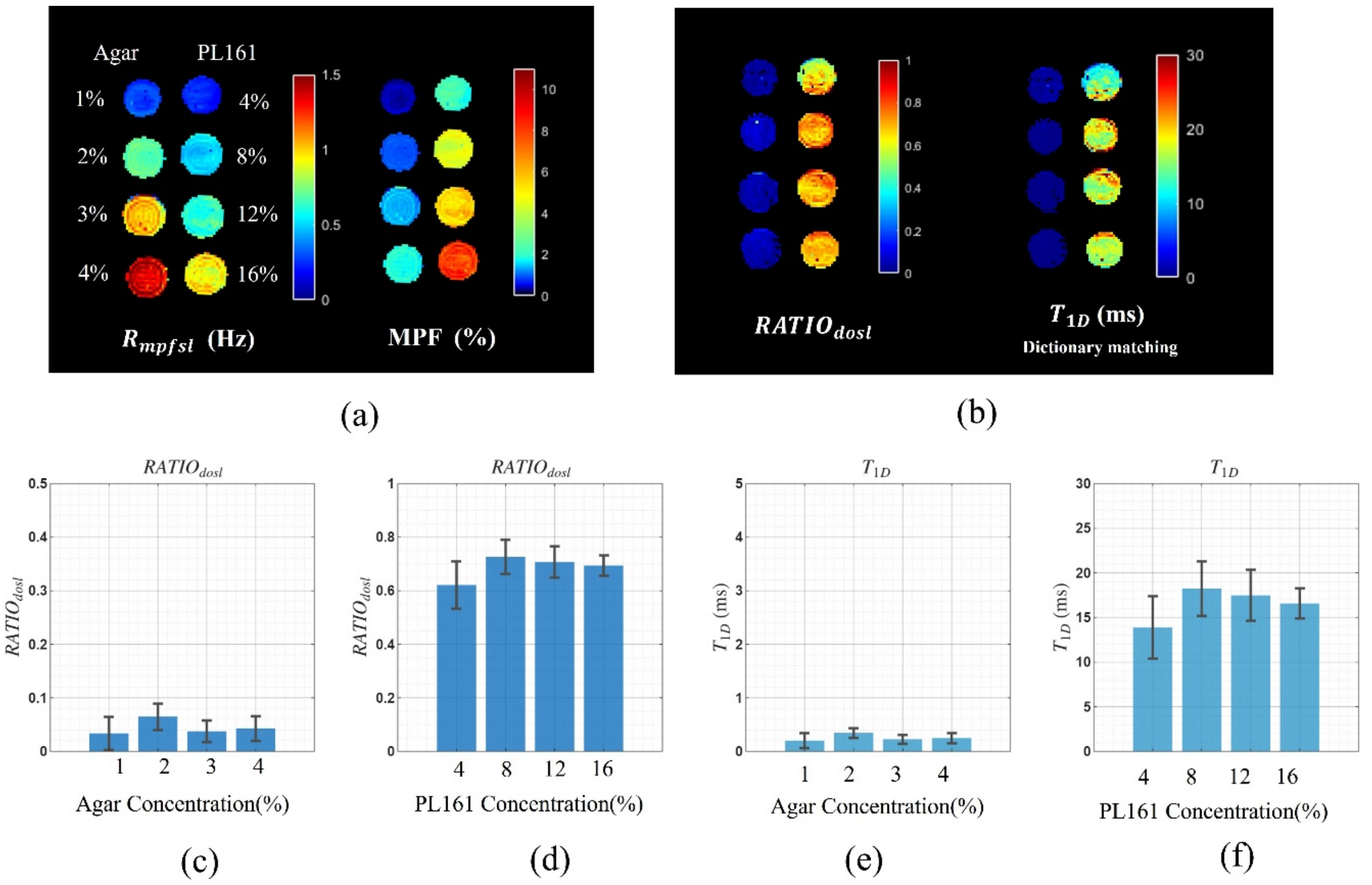

**Figure 7.** Results of phantom studies. (a) $R_{mpfsl}$ and MPF maps. (b) $RATIO_{dosl}$ and $T_{1D}$ map. The first column displays agar phantoms with concentrations of 1%, 2%, 3%, and 4% (from top to bottom); the second column displays PL161 phantoms with concentrations of 4%, 8%, 12%, and 16% (from top to bottom). (c)–(d) Bar plots of $RATIO_{dosl}$ for the agar and PL161 phantoms, respectively. (e)–(f) Corresponding bar plots of $\mathrm{T_{1D}}$.

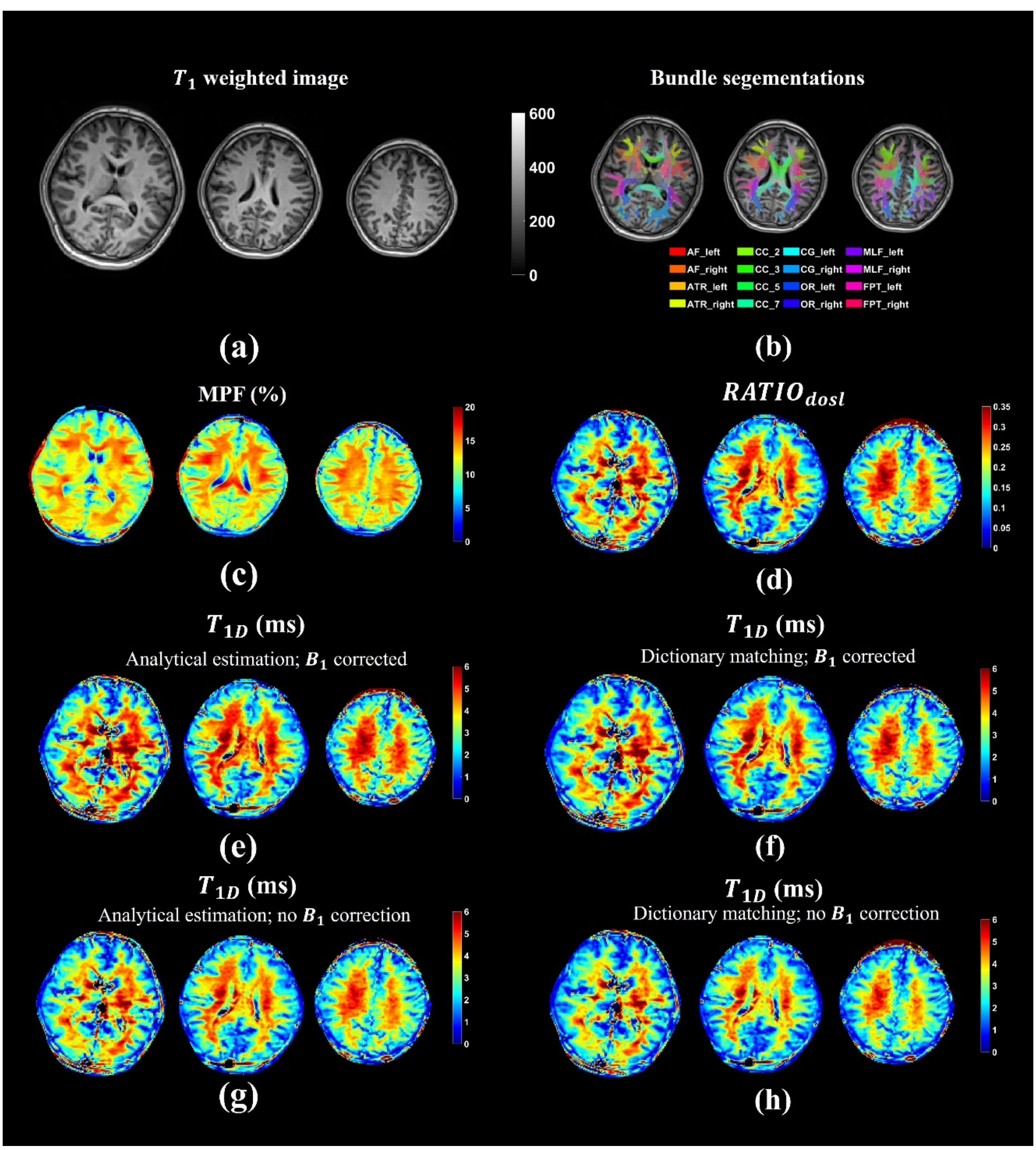

**Figure 8.** Representative results from one volunteer. (a) $T_1$-weighted image of the acquired slices. (b) Bundle segmentation showing 16 major white matter fiber bundles. (c) MPF maps derived from MPF-SL. (d) $RATIO_{dosl}$ maps. (e)–(f) $T_{1D}$ maps obtained via analytical estimation and dictionary matching, respectively, with $B_1$ correction. (g)–(h) Corresponding $T_{1D}$ maps obtained via analytical estimation and dictionary matching without $B_1$ correction.

**Table 2. Mean and standard deviation of MPF, $\boldsymbol{RATIO}_{\mathbf{dosl}}$, and $\boldsymbol{T_{1D}}$ across the 16 major white matter fiber bundles.**

| | MPF (%) | $RATIO_{dosl}$ | $B_1$ correction | | no $B_1$ correction | |
|---|---|---|---|---|---|---|
| | | | $T_{1D}$ (ms) (Analytical estimation) | $T_{1D}$ (ms) (Dictionary matching) | $T_{1D}$ (ms) (Analytical estimation) | $T_{1D}$ (ms) (Dictionary matching) |
| AF_left | 14.51±0.43 | 0.26±0.01 | 4.46±0.17 | 4.17±0.19 | 4.10±0.16 | 3.86±0.17 |
| AF_right | 14.53±0.38 | 0.27±0.01 | 4.55±0.19 | 4.27±0.21 | 4.23±0.17 | 3.99±0.19 |
| ATR_left | 13.60±0.56 | 0.24±0.02 | 4.10±0.30 | 3.81±0.31 | 3.78±0.28 | 3.53±0.28 |
| ATR_right | 13.69±0.36 | 0.24±0.02 | 4.00±0.29 | 3.69±0.28 | 3.74±0.26 | 3.47±0.26 |
| CC_2 | 13.93±0.42 | 0.26±0.01 | 4.34±0.17 | 4.07±0.20 | 4.04±0.16 | 3.79±0.18 |
| CC_3 | 13.98±0.49 | 0.28±0.02 | 4.53±0.25 | 4.42±0.32 | 4.30±0.27 | 4.08±0.29 |
| CC_5 | 14.04±0.59 | 0.28±0.02 | 4.74±0.29 | 4.52±0.34 | 4.36±0.25 | 4.15±0.28 |
| CC_7 | 13.66±0.43 | 0.26±0.01 | 4.44±0.19 | 4.14±0.21 | 4.03±0.17 | 3.77±0.19 |
| CG_left | 14.18±0.38 | 0.27±0.01 | 4.59±0.18 | 4.30±0.20 | 4.21±0.16 | 3.97±0.18 |
| CG_right | 13.84±0.33 | 0.24±0.01 | 4.11±0.15 | 3.79±0.16 | 3.79±0.14 | 3.52±0.15 |
| OR_left | 13.47±0.49 | 0.24±0.03 | 4.25±0.44 | 3.94±0.43 | 3.84±0.40 | 3.59±0.39 |
| OR_right | 13.52±0.35 | 0.25±0.01 | 4.38±0.19 | 4.07±0.20 | 3.98±0.18 | 3.73±0.19 |
| MLF_left | 14.14±0.45 | 0.27±0.01 | 4.56±0.17 | 4.26±0.19 | 4.16±0.16 | 3.91±0.17 |
| MLF_right | 14.11±0.43 | 0.29±0.01 | 4.82±0.18 | 4.59±0.22 | 4.44±0.18 | 4.22±0.20 |
| FPT_left | 14.37±0.40 | 0.30±0.01 | 5.02±0.22 | 4.80±0.26 | 4.64±0.20 | 4.45±0.23 |
| FPT_right | 14.29±0.35 | 0.28±0.01 | 4.65±0.20 | 4.40±0.23 | 4.33±0.18 | 4.11±0.20 |